\documentclass[12pt]{iopart}

\usepackage{lineno,hyperref}
 \usepackage{color}
 \usepackage{bm}
 \usepackage[outdir=./]{epstopdf}
 \usepackage{graphicx, subfigure}
 \usepackage{adjustbox}
 \usepackage{latexsym}
 \usepackage{pifont}
 \usepackage{amssymb}
 \usepackage{hyperref}
 \usepackage{titlesec}
 \expandafter\let\csname equation*\endcsname\relax
 \expandafter\let\csname endequation*\endcsname\relax
 \usepackage{amsmath}
 \usepackage{listings}
 \usepackage{verbatim}
 \usepackage{geometry}
 \usepackage{booktabs}
 \usepackage{enumitem}

\modulolinenumbers[5]

 \definecolor{red}{rgb}{1.0,0.0,0.0}
 \definecolor{gre}{rgb}{0.0,1.0,0.0}
 \definecolor{blu}{rgb}{0.0,0.0,1.0}

 \definecolor{ora}{rgb}{1.0,0.5,0.0}
 \definecolor{gra}{rgb}{0.0,0.5,1.0}

 \newcommand{\ba}{\begin{abstract}}
 \newcommand{\ea}{\end{abstract}}
 \newcommand{\be}{\begin{equation}}
 \newcommand{\ee}{\end{equation}}
 \newcommand{\rom}[1]{\uppercase\expandafter{\romannumeral #1 \relax}}
 \newcommand{\Poincare}{Poincar$\acute{\rm e}$ }
 \newcommand{\Fig}[1]{Fig.\ref{fig:#1}}
 
 \newcommand{\Eqn}[1]{Eq.(\ref{eq:#1})}
 
 \renewcommand{\t}{\theta}
 \newcommand{\z}{\zeta}
 \newcommand{\ds}{\displaystyle}
 \newcommand{\vect}[1]{\mathbf{#1}}
 \newcommand{\dd}[1]{d #1}
 \newcommand{\pdv}[2]{\frac{\partial #1}{\partial #2}}
 \newcommand{\qty}{}
 \newcommand{\abs}[1]{|#1|}

 \graphicspath{{./plots/}}
\begin{document}

\title{New method to design stellarator coils without the winding surface}

\author{Caoxiang Zhu$^1$, Stuart R. Hudson$^2$, Yuntao Song$^3$, Yuanxi Wan$^1$}
\address{$^1$ University of Science and Technology of China, 96 JinZhai Road, Hefei, Anhui 230026, P. R. China}
\address{$^2$ Princeton Plasma Physics Laboratory, Princeton University, P.O. Box 451, New Jersey 08543, USA}
\address{$^3$ Institute of Plasma Physics, Chinese Academy of Sciences, Hefei, Anhui 230031, P. R. China}
\ead{zcxiang@mail.ustc.edu.cn}

\begin{abstract}
Finding an easy-to-build coils set has been a critical issue for stellarator design for decades.
Conventional approaches assume a toroidal ``winding" surface.
We'll investigate if the existence of winding surface unnecessarily constrains the optimization, and a new method to design coils for stellarators is presented.
Each discrete coil is represented as an arbitrary, closed, one-dimensional curve embedded in three-dimensional space.
A target function to be minimized that covers both physical requirements and engineering constraints is constructed.
The derivatives of the target function are calculated analytically.
A numerical code, named FOCUS, has been developed.
Applications to a simple configuration, the W7-X, and LHD plasmas are presented.
\end{abstract}

\section{Introduction}

The difficulties in designing and fabricating current-carrying coils to produce the magnetic fields required for confining plasmas for the purpose of creating fusion energy has been a critical problem since the beginning of research into magnetically confined plasmas in the 1950s.
This problem remains crucial, as is evidenced by the recent NCSX \cite{lessonncsx} stellarator, which was partially built at Princeton Plasma Physics Laboratory but, because of budget constraints, was ultimately cancelled; and by the construction delays of the W7-X experiment \cite{w7xdelay} recently completed in Germany.
The construction of the coils is only one component of modern experiments; but, realizing that it is the currents in the coils that produce the ``magnetic bottle'' that confines the plasma, it is easy to understand that designing and accurately constructing suitable coils is paramount.

Tremendous efforts have been made to find optimal coils that meet both the ``physics'' requirements of producing the desired magnetic field as precisely as possible; and the ``engineering'' requirement that it is realistically possible to construct the coils using existing technology.
Coils that are well-separated, with ample space for the vacuum vessel and other components of the experiment, and with simple shapes are easier and {\em cheaper} to build.
 The task of identifying suitable coil configurations has a direct impact of the cost of magnetic confinement experiments, and thus also on fusion research, and indeed ultimately on the cost of the electricity \cite{ariescs_coils}. 

Early stellarator designs, which pre-date the advent of modern computational architectures, usually began with the coils being described analytically, like the general ``winding law" \cite{windinglaw} by which the current filaments on a given toroidal surface are represented using trigonometric functions. 
By varying the coil parameters, the geometry of the coils can be varied to simultaneously obtain optimal plasma properties and to satisfy engineering constraints. 
However, the winding law represents a very restricted class of coil configurations. 
 
Another approach, introduced by N\"uhrenberg \& Zille \cite{StellaratorOptimize}, allowed the development of a ``fully-optimized" stellarator by separating the problem of designing a high-performance plasma from the task of designing a suitable coils set.
First, an attractive plasma equilibrium is identified by using a nonlinear optimization, where the equilibrium is typically calculated \cite{Bauer1984} using the macroscopic model of ideal magnetohydrodynamics, for which the geometry of the plasma boundary, ${\cal S}$, and various profiles such as the assumed pressure and current are required as boundary conditions.
These boundary conditions are adjusted to discover an equilibrium that is stable with respect to small external perturbations and internal oscillations, reduces the transport and turbulent properties, and so forth.

The second stage of the design, which is the topic of this article, is to design a set of discrete current-carrying coils that creates the required external ``vacuum'' field for confining the ``reference plasma'' configuration.
Hereafter, we will assume that a desired, reference plasma configuration is provided. 
 
The vacuum field, ${\bf B}_V$, by which we mean the magnetic field produced by currents external to the plasma domain, must balance the magnetic field, ${\bf B}_P$, produced by the currents that may or may not be present in the plasma, so that the normal component of the {\em total} magnetic field, ${\bf B} = {\bf B}_V + {\bf B}_P$, on ${\cal S}$ is zero.

For vacuum fields, the Ampere's Law in ideal magnetohydrodynamics (MHD) equations reduces to $\nabla \times {\bf B}_V=0$.
Together with the magnetic divergence constraint $\nabla \cdot {\bf B}_V=0$, we can easily derive $\nabla^2 \phi = 0$, where $\phi$ is the magnetic scalar potential.
Thus, the coil determination problem is to solve Laplace's equation with the boundary condition of $\vect{B}_V \cdot \vect{n} = - \vect{B}_P \cdot \vect{n}$ on ${\cal S}$ ($\vect{n}$ is the unit surface normal).
The plasma boundary is usually specified by Fourier harmonics $R_{mn}, Z_{mn}$ over the poloidal angle $\t$ and toroidal angle $\z$ (restricting attention to stellarator symmetry \cite{stellaratorsymmetry} for convenience),
 \begin{align}
 \ds R &= \sum R_{mn} \, \cos(m\t - n\z) \ ,\nonumber \\
 \ds Z &= \sum Z_{mn} \, \sin(m\t - n\z) \ . \nonumber
 \end{align}

Pioneering work in the field of coil design was performed by Merkel with the development of NESCOIL code \cite{nescoil}, in which he assumed that the external magnetic field is produced by a surface current distribution on a closed toroidal surface surrounding the plasma.
This toroidal surface constrains the location of the coils and is called the ``current carrying surface" or the ``winding surface". 
The surface current density is expressed by a current potential $\Phi$ on the winding surface, $\vect{j} = \vect{n} \times \nabla \Phi$, where ${\bf n}$ is normal to the winding surface.
A Green's function method is then applied to solve the current potential distribution which minimizes the squared normal error $\ds \epsilon^2 = \oint_{\cal S} ( \vect{B} \cdot \vect{n} )^2 \dd{s}$.
Once the surface current potential is determined, a set of discretized coils can be obtained by selecting an appropriate number of contours of $\Phi$.
This method was successfully applied to the design of the coils of W7-X \cite{w7x}, the world's largest ever stellarator.
 
Merkel's method leads to a Neumann condition problem that can be linearly solved. 
So it's inherently fast and robust. 
However, due to the ill-conditioned nature of inverting the Biot-Savart integrals, the NESCOIL results for plasmas with complex shapes are usually not viable (e.g. unreasonably large amplitudes in high modes or impractically complex curvatures). 
A singular value decomposition (SVD) method \cite{pomphrey2001} and, more recently, a Tikhonov regularization approach \cite{REGCOIL} were applied to provide improvements on NESCOIL.
Nevertheless, these methods only indirectly control the geometry of the resultant coils set, and allow limited opportunities to optimize the engineering constraints.
 
A different approach that explicitly incorporates engineering constraints  has been advanced by Drevlak with the extended NESCOIL code \cite{drevlak1998} and ONSET \cite{ONSET}, by Strickler {\em et al.} with the code COILOPT \cite{coilopt01}, and later by Breslau {\em et al.} with COILOPT++ \cite{coilopt++}.
The coils are represented as ``filaments'', one-dimensional curves, lying on a toroidal winding surface (pre-defined or optimized simultaneously). 
The magnetic field produced by $\delta$-function current-densities in the coils set is calculated using the Biot-Savart law.
The geometry of the coils is varied using nonlinear optimization algorithms to minimize a ``cost-function'', which represents a balance between the physics requirements (that the total normal magnetic field at the plasma boundary is as small as possible) and the engineering constraints (that the coils can be achieved by current engineering techniques). 
Coils of the compact stellarators NCSX \cite{NCSX} and ARIES-CS \cite{ariescs_lpku} are designed in such approaches.
 
For all the methods mentioned above, a toroidal winding surface is required to locate the coils (ONSET may need two constraining surfaces for interpolation).
For NESCOIL and NESCOIL-like codes (NESVD, REGCOIL, etc.), the winding surface is strictly required, as it provides the surface where the current potential lies so that the currents can be explicitly represented and then be solved.
For the discrete coils nonlinear optimization codes, the winding surface is not strictly required, but there are advantages.
This decreases the number of freedoms and the plasma-coil separation, which is important for divertor/blanket access, can be easily enforced by the existence of winding surface.
 
However, on the other hand, the existence of winding surface also provides strong limitations.
A ``bad" winding surface directly results in the failure of finding acceptable coils set.
Thus, before getting an acceptable coils set, a ``good" winding surface should be obtained first. 
However, the standard of good or bad of a winding surface is actually evaluated by the final coil shapes.
So nested optimizations targeting both the winding surfaces and the coils are required to be carried out back and forth.
 
In this paper, we present a new method for designing stellarator coils that eliminates the need of a toroidal winding surface.
The new coil representation, as well as the interpretations on the definition of objective functions and their analytical derivatives, are discussed in section \rom{2}.
The numerical implementation, a code named Flexible Optimized Coils Using Space curves (FOCUS), is introduced in section \rom{3}.
In section \rom{4}, we show some example calculations.
Conclusions and future works are discussed in section \rom{5}. 

\section{3D coil representations}
\subsection{The Fundamental Theorem of Curves}
For convenience, we work within the filamentary approximation, by which we mean that each coil has zero cross-sectional area.
 (This approximation is not fundamental, and can easily be relaxed in future work.)
We represent the geometry of the coils as closed, one-dimensional curves embedded in three-dimensional space.
 
The geometry of such curves is elegantly described by the Fundamental Theorem of Curves \cite{diffgeom}, which states that any regular curves in three-dimensional space with non-zero curvature can be determined, up to rigid displacements and rotations, by its curvature, $\kappa(s)$, and torsion, $\tau(s)$, where $s$ parameterizes arc-length.
Given $\kappa$ and $\tau$, and boundary conditions (such as the starting point), the geometry, ${\bf x(s)}$, unit tangent vector, ${\bf t(s)}$, unit normal and bi-normal, ${\bf n(s)}$ and ${\bf b(s)}$, of the curve can be obtained by integration of the Frenet-Serret formulas \cite{FrenetSerret},
 \begin{align} \label{eq:curve}
 \begin{bmatrix}
 \vect{x} \\
 \vect{t} \\
 \vect{n} \\
 \vect{b} \\
 \end{bmatrix}'
 =
 \begin{bmatrix}
 1 &      0 & 0 \\
 0 & \kappa & 0 \\
 - \kappa & 0 & \tau \\
 0 & \tau & 0 \\
 \end{bmatrix}
 \ 
 \begin{bmatrix}
 \vect{t} \\
 \vect{n} \\
 \vect{b} \\
 \end{bmatrix},
 \end{align}
 where the ``prime'' denotes derivative with respect to the parameter $s$, i.e., ${\bf x}' \equiv d{\bf x} / ds$.
 
The benefit is that each curve is described by two independent functions of arc length, namely the curvature and torsion. 
Using the curvature function as an independent degree of freedom to implicitly define the coil geometry has its appeal, as the curvature of the coils is frequently an engineering constraint and any curvature constraints can potentially be enforced before the nonlinear optimization proceeds.
However, solving 12 ODEs for each coil at each iteration of the nonlinear optimization turns out to be somewhat cumbersome.
Furthermore, to exploit efficient nonlinear optimization algorithms, we shall later construct the derivatives of the coil geometry with respect to the independent degrees of freedom, and this requires integration of the derivatives of \Eqn{curve} with respect to the curvature and torsion, which becomes even more cumbersome.
Besides, additional constraints, such as ensuring the curves closed and sufficiently smooth, must also be included.
 
\subsection{Fourier representation}
Instead, we employ an easier way of describing one-dimensional curves embedded in three-dimensional space.
A curve is described directly, and completely generally, in Cartesian coordinates as $\vect{x}(t) = x(t) \, {\bf i} + y(t) \, {\bf j} + z(t) \, {\bf k}$.
Three functions are required to specify the geometry, namely $x(t)$, $y(t)$ and $z(t)$, with the constraints that each function be periodic, e.g. ${\bf x}(t+T) = {\bf x}(t)$ for some $T$.
The curve parameter, $t$, at this point is arbitrary.
 
A variety of mathematical representations are possible.
For purpose of illustration, and because our initial interest is in smooth coils, we use a Fourier representation:
\begin{align} \label{eq:fseries}
x(t) & =  X_{c,0} + \sum_ {n=1}^{N_F} \left [X_{c,n} \cos(nt) + X_{s,n} \sin(nt) \right ], 
\end{align}
with $t$ varying between $[0, 2\pi]$, and similarly for $y(t)$ and $z(t)$.
The shape of a coil is then fully determined by the $3 \times (2N_F+1)$ Fourier coefficients.
No additional assumptions are made here, so this representation fits for all kinds of smooth coils, such as helical, modular, saddle, etc.
(For coils with straight segments, for example the rectangular and ``window pane'' coils, the Fourier representation is not suitable.
In future upgrades to FOCUS, additional representations for the curves, such as cubic splines and piecewise-linear, will be implemented.)
 
Hereafter, $\vect{X}$ will be used to represent the geometrical degrees of freedom of a set of $N_C$ coils, ${\bf x}_i$, and the coil currents, $I_i$, which are also considered as independent degree of freedom.
The total number of degrees of freedom is $N_{D} = N_C \times [ 3 \times (2N_F+1) + 1 ]$.

\section{Theory of the coil optimization problem}
  
\subsection{General objective functions}
The coil parameters are to be varied to minimize a target function consisting of both ``physics'' and ``engineering'' objective functions,
\begin{equation} \label{eq:target}
\ds \chi^2(\vect{X}) = \sum_j w_j \left( \frac{f_j({\vect X}) - f_{j,o}}{f_{j,o}} \right)^2,
\end{equation}
where $f_j({\vect X})$ is the value of the $j^{th}$ objective function, to be defined below, for a given set of coil parameters, $f_{j,o}$ denotes the desired value, and $w_j$ is a user-prescribed weight.

 \subsection{Normal field error on plasma boundary}
 The key criteria is to satisfy the condition of $B_n = \vect{B} \cdot \vect{n} = 0$ on the pre-defined plasma boundary as much as possible with a finite number of coils.
  A smaller $B_n$ error usually implies a better approximation to the target equilibrium properties.
 This can be achieved by minimizing the squared relative error of the total normal field integrated over the boundary,
 \begin{equation} \label{eq:bnormal}
 \ds f_B(\vect{X}) \equiv \int_S \frac{1}{2} \left ( \frac{\vect{B} \cdot \vect{n}}{\left | \vect{B} \right |} \right )^2 \dd{s}.
 \end{equation}

 The magnetic field at position $\vect{\bar x}$ produced by a coils set is calculated using the Biot-Savart law,
 \be \label{eq:biotsavart}
 \ds \vect{B}_V({\bf \bar x}) = \frac{\mu_0}{4\pi}\sum_{i=1}^{N_C} I_i \ \int_{C_i} \frac{\dd{\vect{l}_i} \times \vect{r} }{r^3}.
 \ee
 Here, $I_i$ is the current in $i^{th}$ coil, $\dd{\vect{l}_i} \equiv {\bf x}_i' \, dt$ is the differential line element, and $\vect{r} = \vect{\bar x}- \vect{x}_i$ is the displacement vector between the evaluate point on the surface and the differential element.
 
 The plasma field, $\vect{B}_{P}$, produced by the reference plasma must be pre-computed using a suitable equilibrium code, and the following will assume that the plasma field does not change during the optimization of the coil geometry.
The FOCUS code can accommodate an arbitrary $\vect{B}_{P}$, but for simplicity of illustration in this paper we restrict attention to vacuum fields, so that $\vect{B} = \vect{B}_{V}$, and we hereafter suppress the subscript $V$ for notational clarity.

 If a small geometry deformation $\delta \vect{x}_i$ is applied to the $i^{th}$ coil, the corresponding change in $\vect{r}$ and $r$ are $\delta \vect{r} = - \delta \vect{x}_i$ and $\delta r = - \vect{r} \cdot {\delta \vect{x}_i}/r$, and the change in the magnetic field is
 \be
 \ds \delta \vect{B}({\bf \bar x}) = \frac{\mu_0}{4\pi} I_i \int_0^{2\pi} \left [ \frac{3\vect{r} \cdot \vect{x}_i'}{r^5}  \vect{r}\times \delta \vect{x}_i  \, + \, \frac{2}{r^3}\delta \vect{x}_i  \times \vect{x}_i'  \, + \, 
 \frac{3\vect{r} \cdot \delta \vect{x}_i}{r^5} \vect{x}_i'  \times \vect{r} \right ] \dd{t}\;  . 
 \ee

 The change in $f_B$ is, omitting the normalization term for notational clarity, given by
 \be \label{eq:Obderiv}
 \ds \delta f_B(\vect{X}) = \int_S (\vect{B} \cdot \vect{n}) \, \frac{\mu_0}{4\pi} I_i \int_0^{2\pi} \left [ \frac{3\vect{r} \cdot \vect{x}_i'}{r^5} \vect{n} \times \vect{r} \, + \, 
 \frac{2}{r^3} \vect{x}_i' \times \vect{n}  \, + \, \frac{3 \vect{x}_i' \times \vect{r} \cdot \vect{n}}{r^5} \vect{r}\right ] \cdot \delta \vect{x}_i \; \dd{t} \;
\dd{s}. 
\ee

What should be noted here is that the variation of $f_B$ (as well as other object functions) with respect to coil currents $\delta I_i$ can be derived in the same way. But we will not cover the details it in this article.

 \subsection{Constraint of the toroidal flux}
 To avoid trivial solutions, like when $I_i \rightarrow 0$ $\forall i$, $f_B \rightarrow 0$, it is sufficient to constrain the enclosed toroidal flux.
 If ${\bf B}\cdot{\bf n}=0$ on the boundary, then the toroidal flux through any poloidal cross-sectional surfaces is constant.
 We include an objective function defined as
 \be
 \ds f_\Psi(\vect{X}) \equiv \frac{1}{2\pi} \int_0^{2\pi} \frac{1}{2} \left( \frac{\Psi_\z \; - \; \Psi_o}{\Psi_o} \right)^2 \dd{\z},
 \ee
 where the flux through a poloidal surface, ${\cal T}$, produced by cutting the boundary with plane $\z=const.$ is computed using Stokes' theorem,
 \begin{align}
 \ds \Psi_\z({\bf X}) & \equiv \int_{\cal T} \! \vect{B} \cdot \dd{\vect{S}} = \oint_{\partial {\cal T}} \!\! \vect{A} \cdot \dd{\vect{l}}.
 \end{align}
 Here $\dd \vect{l}$ is on the boundary curve of the poloidal surface and the total magnetic vector potential $\vect{A}$ is 
 \begin{align} \vect{A}({\bf \bar x})  = \frac{\mu_0}{4\pi}\sum_{i=1}^{N_C} I_i \ \int_{C_i} \frac{\dd{\vect{l}_i}}{r}.
 \end{align}
 The variation of $f_\Psi$ resulting from $\delta \vect{x_i}$ is 
 \begin{align}\label{eq:Otderiv}
 \ds \delta f_\Psi(\vect{X}) &= \frac{1}{2\pi} \int_0^{2\pi} \left (\frac{\Psi_\z \; - \; \Psi_o}{\Psi_o} \right ) \ \frac{\delta \Psi_\z}{\Psi_o}\dd{\z} ,
 \end{align}
 where
 $\ds \delta \Psi_\z = \int_{\partial {\cal T}} \delta \vect{A} \cdot \dd{\vect{l}}$ and 
 \begin{align} 
 \ds \delta \vect{A}({\bf \bar x}) &= \frac{\mu_0}{4\pi} I_i \ \int_0^{2\pi} \left [ -\frac{\vect{r} \cdot \vect{x}_i'}{r^3} \ \delta \vect{x}_i \, + \, \frac{\vect{r} \cdot \delta \vect{x}_i}{r^3} \ \vect{x}_i' \right ] \dd{t} .
 \end{align}
 
\subsection{Engineering constraints}
Besides the physical requirements, engineering constraints should be considered, such as the coil-coil separation, coil-plasma separation, minimum radius of curvature, electromagnetic forces, stored energy, etc.
  
Some of the engineering constraints are actually consistent with the magnetic field requirements.
For example, to reduce the magnetic field ripple, the coils are usually placed away from the plasma, which is exactly of the same effect as maximizing the coil-plasma separation constraint for the capacity of divertors.
And the coil-coil separation constraint is to prevent coils overlapping with each other. 
It also coincides with the need of lowering magnetic ripple.
With finite number of coils, it's better to equally place the coils in the toroidal angle; otherwise, the ripple would increase rapidly.

 There are also engineering considerations that compete with the magnetic field requirement.
 Among them, the coil length constraint is the most straightforward.
 Without a constraint on length, the coils can become arbitrarily long to lower the ripple, and more ``wiggles" can potentially be formed to better match the plasma shape.
 Besides, the total coil length is directly related to the usage of materials, i.e. cost, which is without any relation to the physical requirements.
 We include an objective function of the form
 \be
 \ds f_L(\vect{X}) = \frac{1}{N_C} \; \sum_{i=1}^{N_C} \frac{e^{L_i}}{e^{L_{i,o}}},
 \ee
 where $L_i({\bf X})$ is the length of $i$-th coil, 
 \be
 \ds L_i(\vect{X}) = \int_0^{2\pi} \!\! \abs{\vect{x}_i'} \; \dd{t},
 \ee 
 and $L_{i,0}$ is a user-specified normalization.
 The variations in $f_L$ resulting from a variation $\delta \vect{x}_i$ is
 \be\label{eq:Olderiv}
 \ds \delta f_L(\vect{X})  = \frac{1}{N_C} \; \frac{e^{L_i}}{e^{L_{i,0}}}  \int_0^{2\pi} \frac{({\vect{x}'_i} \cdot {\vect{x}'_i}){\vect{x}''_i} - ({\vect{x}'_i} \cdot {\vect{x}''_i}){\vect{x}'_i}}{({\vect{x}_i'} \cdot {\vect{x}_i'})^{3/2}} \cdot \delta \vect{x}_i \ \dd{t}. 
 \ee

\subsection{Spectral condensation} 
The parameterization given in \Eqn{fseries} is not unique.
Consider the curve variation \mbox{$\delta {\bf x}_i = {\bf x}'_i(t) \delta u$}, where $\delta u$ is a small constant.
Inserting this into \Eqn{Obderiv}, \Eqn{Otderiv} and \Eqn{Olderiv}, we can easily get that $\delta f_B$, $\delta f_\Psi$ and $\delta f_L$ are all zero.
This ``variation'' is tangential to the curve and leaves the geometry of curve unchanged, but it does effectively change the parameterization of the curve and thereby changes the Fourier harmonics.
 
This non-uniqueness of the curve parameters can be exploited for numerical efficiency.
For each coil, a spectral width of the Fourier harmonics is defined as,  \cite{spectralcondensation}
 \begin{align}
 \ds M_i(\vect{X}) = \sum_ {n=1}^{N_F} n^p({X^i_{c,n}}^2 +{X^i_{s,n}}^2 +{Y^i_{c,n}}^2 +{Y^i_{s,n}}^2 +{Z^i_{c,n}}^2 +{Z^i_{s,n}}^2),
 \end{align}
where $p \geq 1$. 
Curve parameterizations that minimize $M=\sum_i^{N_C} M_i$ are attractive for numerical purposes, but minimizing the spectral width $M$ should not compete with the physics or engineering properties. 
So when minimizing $M$, we only allow tangential variations in the geometry of coils.
In the Fourier parameter space, the tangential variation is defined as $\delta {\bf X} \equiv {\bf X}' \delta u$, where ${\bf X}'$ is the tangential direction in parameter space and $\delta u$ is an arbitrary function.
If we choose $\delta u \equiv - \partial_{\bf X} M \cdot {\bf X}'$, the spectral width will be forced to decrease subject without changing the coil geometry.
This shall be used below.

\section{Numerical implementations}
 
\subsection{Analytically calculated derivatives}
Our target function is now composed of the normal field error, toroidal flux error and the length constraint, and is represented as,
 \begin{equation}
 \ds \chi^2(\vect{X}) = w_B \;f_B(\vect{X}) + w_{\Psi} \;f_\Psi(\vect{X}) + w_L \; f_L(\vect{X}) .
 \end{equation}
 The task of designing optimal coils is now reduced to a single-object minimization problem. 
 A lot of minimization algorithms can be applied here, like the famous Levenberg-Marquardt (LM) method \cite{LM} which was successfully used in COILOPT and COILOPT++.
 However, unlike other nonlinear coil-designing codes where the derivatives are computed numerically, we can offer the analytically calculated derivatives from the \Eqn{Obderiv}, \Eqn{Otderiv} and \Eqn{Olderiv}. 
 
 Calculating the derivatives involves line and surface integrals. 
 Each coil is treated as a single filament with $N_{s}$ segments. And the predefined plasma surface is discretized into $N_{\t} \times N_{\z}$ surface elements. 
 We shall illustrate how we implement the numerical calculations in the code by one example. 
 The derivative of the normal field error $f_B$ with respect to an arbitrary cosine harmonic of the $i^{th}$ coil denoted as $X_{c,n}^i$ is computed as,
  \begin{align}
 \ds \pdv{f_B}{X_{c,n}^i} & = \int_S (\vect{B} \cdot \vect{n})  \; \pdv{\vect{B} \cdot \vect{n}}{X_{c,n}^i} \dd{s} \nonumber \\
 \ds &  \approx \sum_{j=1}^{N_\t} \sum_{k=1}^{N_\z} \left [ \left (B_x n_x + B_y n_y + B_z n_z \right ) \left (\pdv{B_x}{X_{c,n}^i} n_x + \pdv{B_y}{X_{c,n}^i} n_y + \pdv{B_z}{X_{c,n}^i} n_z \right ) \sqrt g \; \right ]_{j+\frac{1}{2},k+\frac{1}{2}} \frac{2\pi}{N_{\t}} \frac{2\pi}{N_{\z}} \ .
 \end{align}
 Here, the normalization term is omitted and the integral is evaluated by the mid-point rule. And the calculation of $\partial{B_x} / \partial{X_{c,n}^i}$ , which involves a line integral, can be approximated like,
 \begin{align}
 \ds \pdv{B_x}{X_{c,n}^i}  & \approx \frac{\mu_0}{4\pi} I_i \ \sum_{m=1}^{N_s} \left . - \frac{3\qty (\Delta y \ {z'_t} - \Delta z \  {y'_t}) \ \Delta x \ \cos(nt)}{r^5} \right |_{m} \frac{2\pi}{N_{s}} \ .
 \end{align}
  Similarly, the derivatives for toroidal flux error and coil length constraint can also be calculated (fast and accurately). 
 
 \subsection{Steepest descent method}
 Analytically calculated derivatives have great advantages in accelerating the computation speed and accessing more applicable algorithms. 
 For the proof of concept, we will apply the steepest descent algorithm here.
 Defining an artificial ``time" $\tau$, the descent direction for the DoF vector is given by
 \be \label{eq:descent}
 \ds \pdv{\vect{X}}{\tau} \equiv - \pdv{\chi^2}{\bf X} - {\bf X}' \; \left( \pdv{M}{\bf X} \cdot {\bf X}' \right) \ .
 \ee
 Here, in the RHS, the first term is the usual gradient descent and the second is defined in tangential direction specialized for the spectral condensation.
 So the relationship between the target function and time is:
 \be
 \ds \pdv{\chi^2}{\tau} = \pdv{\chi^2}{\vect{X}} \cdot \pdv{\vect{X}}{\tau} = - \pdv{\chi^2}{\vect{X}} \cdot \pdv{\chi^2}{\bf X} 
                                      \; - \; \left ( \pdv{\chi^2}{\vect{X}} \cdot {\bf X}' \right ) \; \left( \pdv{M}{\bf X} \cdot {\bf X}' \right) 
                                      =  - \left (\pdv{\chi^2}{\vect{X}} \right)^2 .
 \ee
 The first derivative $\partial \chi^2 / \partial \tau$ is always non-positive.
 After reaching the minima ($\partial \chi^2 / \partial {\bf X} = 0$), the spectral term has the same monotonicity,
 \be
 \ds \pdv{M}{\tau} = \pdv{M}{\vect{X}} \cdot \pdv{\vect{X}}{\tau} = - \left ( \pdv{M}{\bf X} \cdot {\bf X}' \right )^2.
 \ee
By solving \Eqn{descent}, the steepest descent method will keep minimizing the target function, which covers the physical requirements (the squared normal field error and the toroidal flux error) and engineering constraint (the length constraint), as well as the numerical non-uniqueness eliminator (the spectral condensation term) .
The ODEs in \Eqn{descent} are integrated using a fixed order Runge-Kutta method with the subroutine D02BJF in the NAG Library \cite{NAG}.
 
The steepest descent method is {\em continuous}. 
This means that the coils cannot pass through the plasma, as this would result in a singularity in the normal field.
We can define the linking number of each coil and the plasma using the Gauss linking integral \cite{gausslinking} with respect to the coil and the magnetic axis: 
\be
\mathrm{link}(C_i, Plasma) = \frac{1}{4\pi} \ \oint_{C_i} \oint_{axis} \frac{\vect{r}_1 - \vect{r}_2}{ | \vect{r}_1 - \vect{r}_2|^3} \cdot \dd{\vect{r}_1} \times \dd{\vect{r}_2} \ .
\ee
With the steepest descent algorithm, this linking number is implicitly conserved during the optimization. 
 
\section{Applications} 
To demonstrate the feasibility of this method, three cases are investigated. 
A simple rotating elliptical plasma boundary is used for investigating the convergence properties of the code.
Illustrations for the existing stellarators W7-X and LHD are presented to explore the code's capacity of designing various coils configuration for realistic stellarators.
 
\subsection{Rotating ellipse} 
A two-field period, rotating elliptical plasma boundary is specified as,
\begin{align}
R = & 3.0     + 0.3 \cos(\t) \quad\quad\quad\quad\quad\quad\quad\quad    - 0.06 \cos(\t - N_P \z), \nonumber \\
Z = & \quad \ - 0.3 \sin(\t) - 0.06 \sin(- N_P \z)      - 0.06 \sin(\t - N_P \z), \nonumber
\end{align}
where $N_P  = 2$.
As one of the simplest configurations for stellarators, the rotating ellipse can be achieved with conventional helical coils (like the LHD machine \cite{LHD}).
Helical coils, with certain pitch angles, can efficiently produce desired rotational transforms, particularly for classical stellarators with helix structures.
However, the access to the plasma  and ease of maintenance are restricted badly. 
Modular coils, on the other hand, can easily overcome these difficulties and they can closely match macroscopic magnetic parameters, but differences in the local structures of the fields do exist \cite{modularstellarator}.
Here, we'll try to use the FOCUS to design modular coils for it, just for demonstrations. 
 
\begin{figure}[!htb]
 \centering
 \subfigure[] {\label{fig:L2initial}\includegraphics[width=0.48\textwidth]{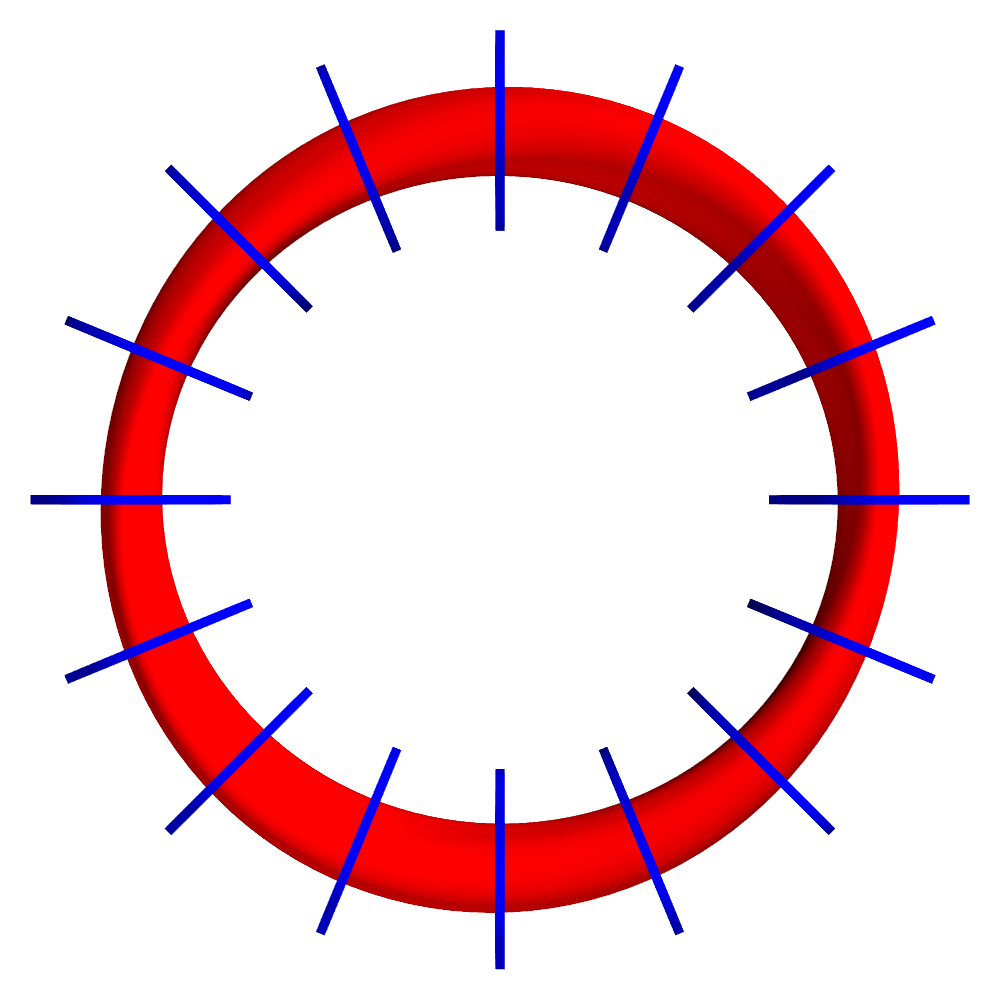}}
 \subfigure[] {\label{fig:L2final}\includegraphics[width=0.48\textwidth]{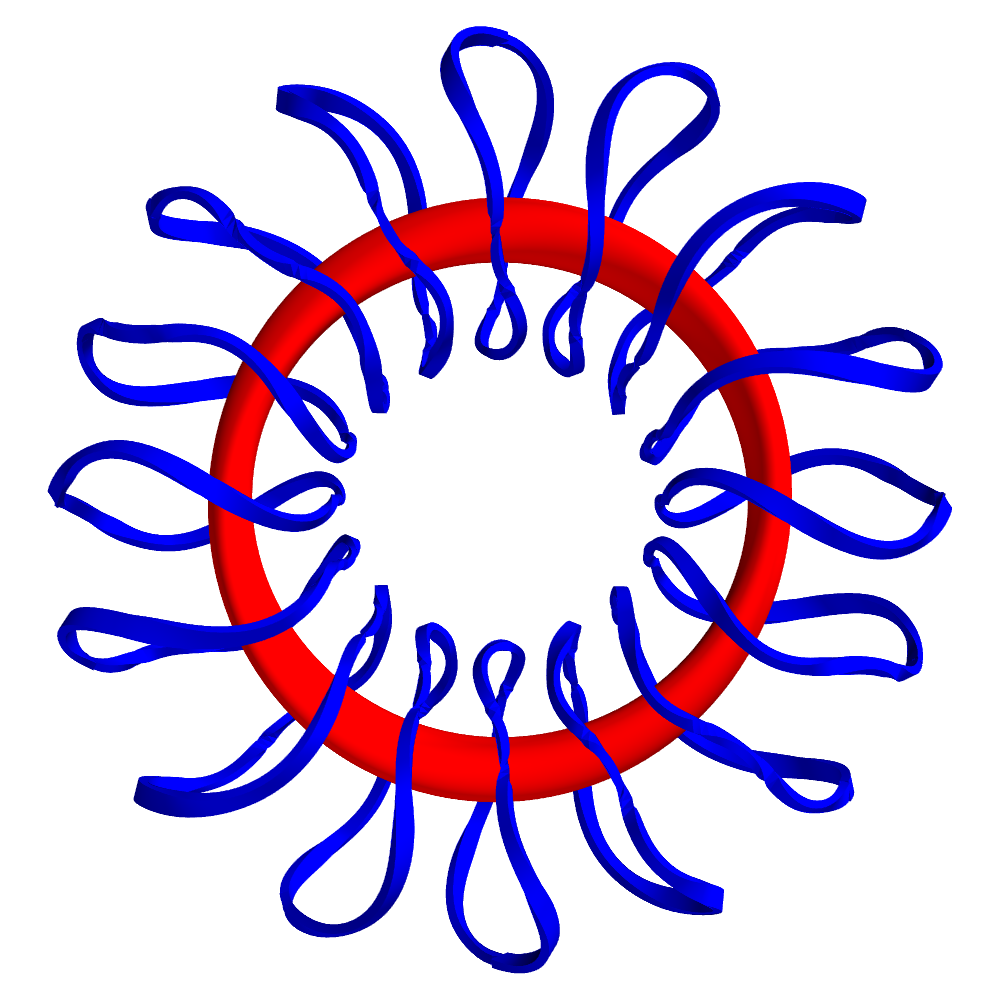}}
 \caption{Coils for the rotating ellipse configuration. (a) The initial 16 circular coils with equal toroidal intervals. (b) The optimized coils with only physical constraints. Red surfaces represents the target plasma boundary. \label{fig:L2coils}}
 \end{figure}

To initialize the coil-design calculation in FOCUS, we must supply a suitable initial guess.
We anticipate that, generally, codes such as NESCOIL and REGCOIL will be used. 
But here, a very simple initial guess for the coils is made by toroidally placing 16 circular coils, as shown in \Fig{L2initial}.
The radius of the coils is $0.75 \ m$, and the chosen Fourier resolution is $N_F=4$.

To test the different objective functions separately, we first include {\em only} the physical constraints, i.e. we set the length constraint weight to zero, $w_L=0$.
The limit on the time integration is set to  an arbitrary value, e.g. $\tau_{end}=10^3$.

\begin{figure}[!htb]
 \centering
 \begin{minipage}[h]{0.48\textwidth}
 \includegraphics[width=\textwidth]{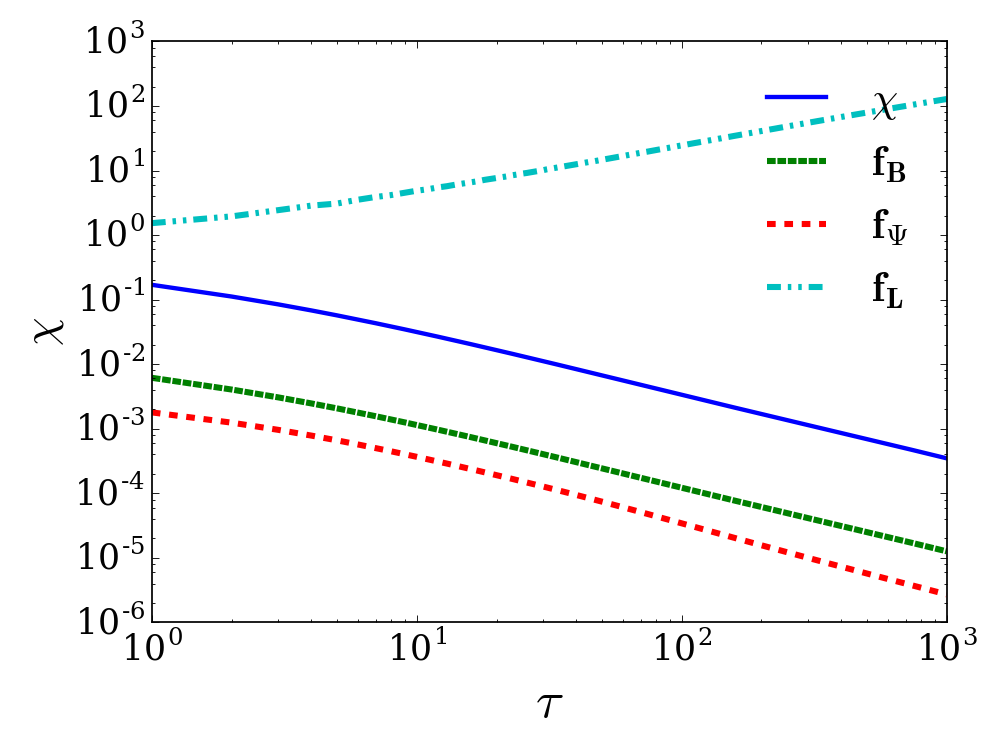}
 \caption{FOCUS optimizing curves of $\chi$, $f_B$, $f_{\Psi}$ and $f_L$ over the integration step $\tau$. \label{fig:L2fit}}
  \end{minipage}
  \hfill
  \begin{minipage}[h]{0.48\textwidth}
 \includegraphics[width=\textwidth]{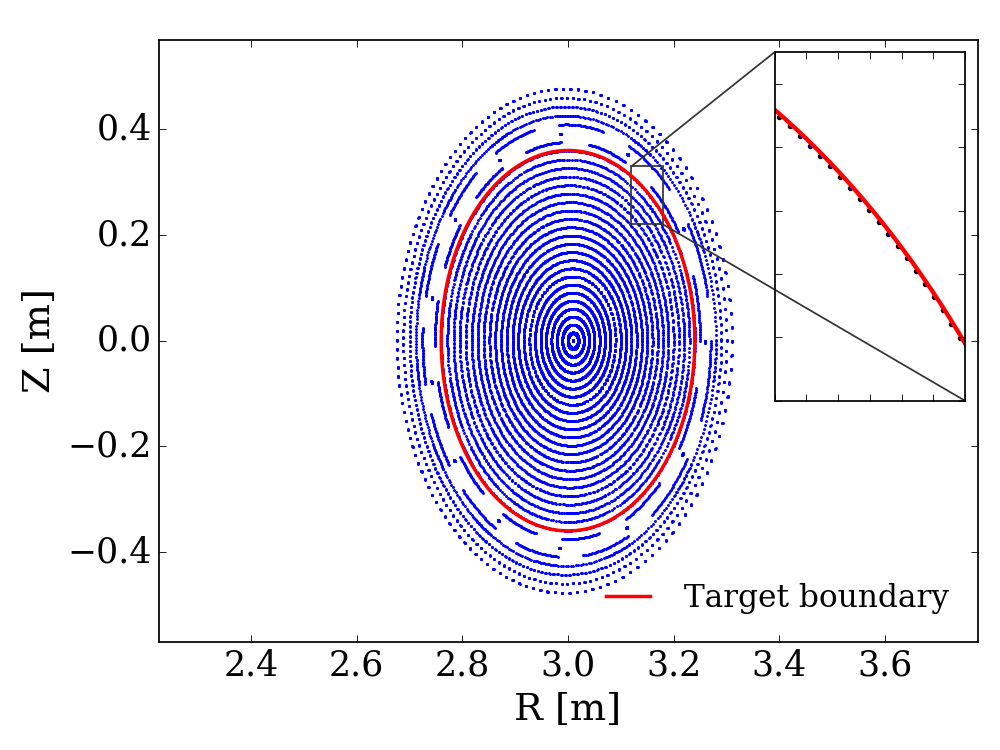}
 \caption{\Poincare plot of the flux surfaces (blue) produced by the optimized coils and the target plasma boundary (red).\label{fig:L2pp}}
 \end{minipage}
 \end{figure}

As illustrated in \Fig{L2fit}, $f_B$ significantly decreases from $3.63\times10^{-2}$ to $1.26\times10^{-5}$ at $\tau = 10^3$.
The final shape of the optimized coils are shown in \Fig{L2final}. 
We can see that all the coils are getting longer and further away from the plasma to better match the plasma and reduce the ripple.
A \Poincare plot of the flux surfaces produced by the optimized coils, \Fig{L2pp}, indicates that a satisfactory magnetic field has been obtained.

Note that in this optimization, the only targets were to minimize $w_B \;f_B + w_{\Psi} \;f_\Psi$ ($w_B / w_{\Psi}=100/1$). Engineering constraints such as coil-coil separation, coil-plasma separation, even the coil length constraint were {\em not} included. 
The coils are not constrained to lie on a winding surface and are completely free to move in space. 
The fact that we get reasonable coils is remarkable.

\Fig{L2fit} also indicates that the target function is not converged at $\tau = 10^3$. 
Actually, it turns out that even at $\tau = 10^6$, the target function is still not converged. 
We'll explore the convergence properties by varying different parameters.

\begin{figure}[!htb]
 \subfigure[] {\label{fig:NFconv}\includegraphics[width=0.48\textwidth]{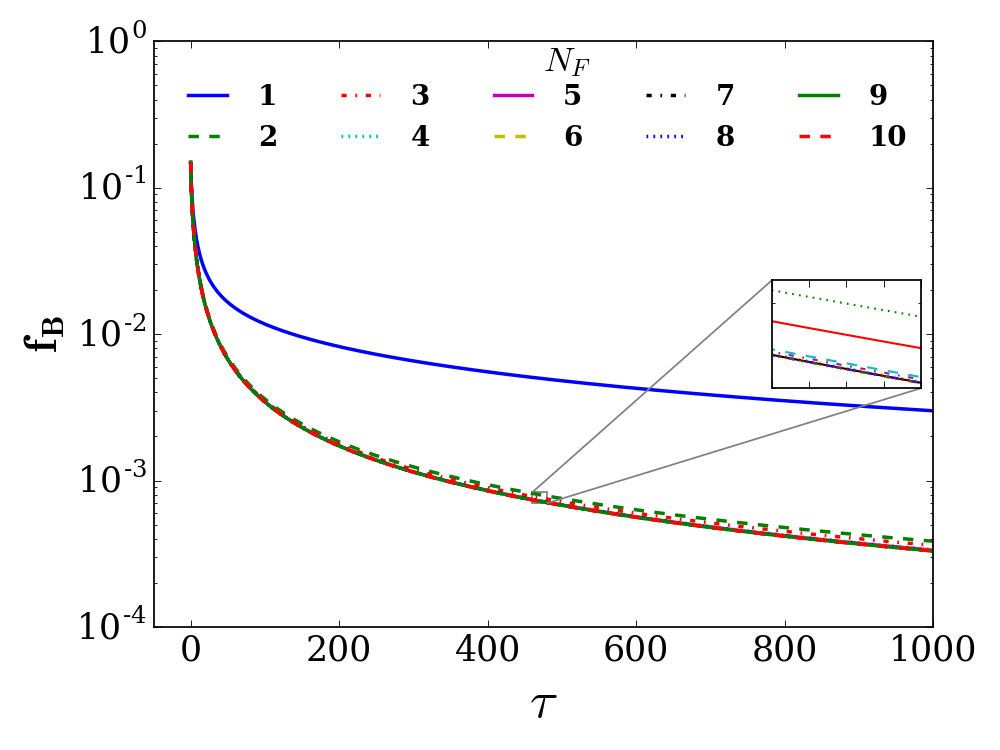}}
 \subfigure[] {\label{fig:Ncconv}\includegraphics[width=0.48\textwidth]{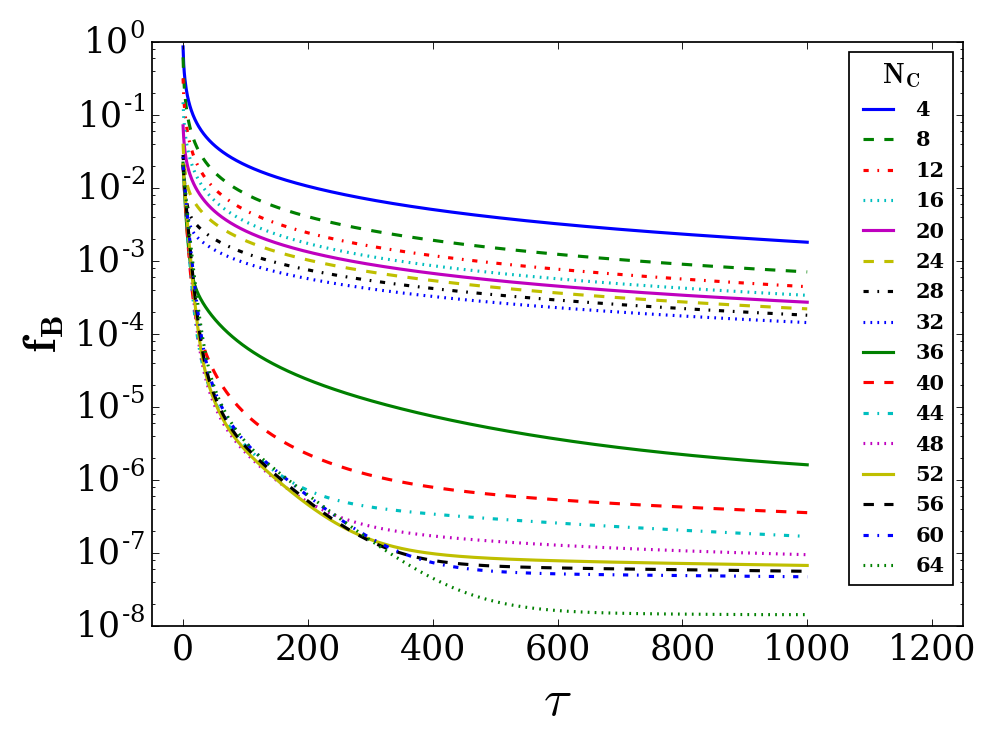}}
 \subfigure[] {\label{fig:Rconv}\includegraphics[width=0.48\textwidth]{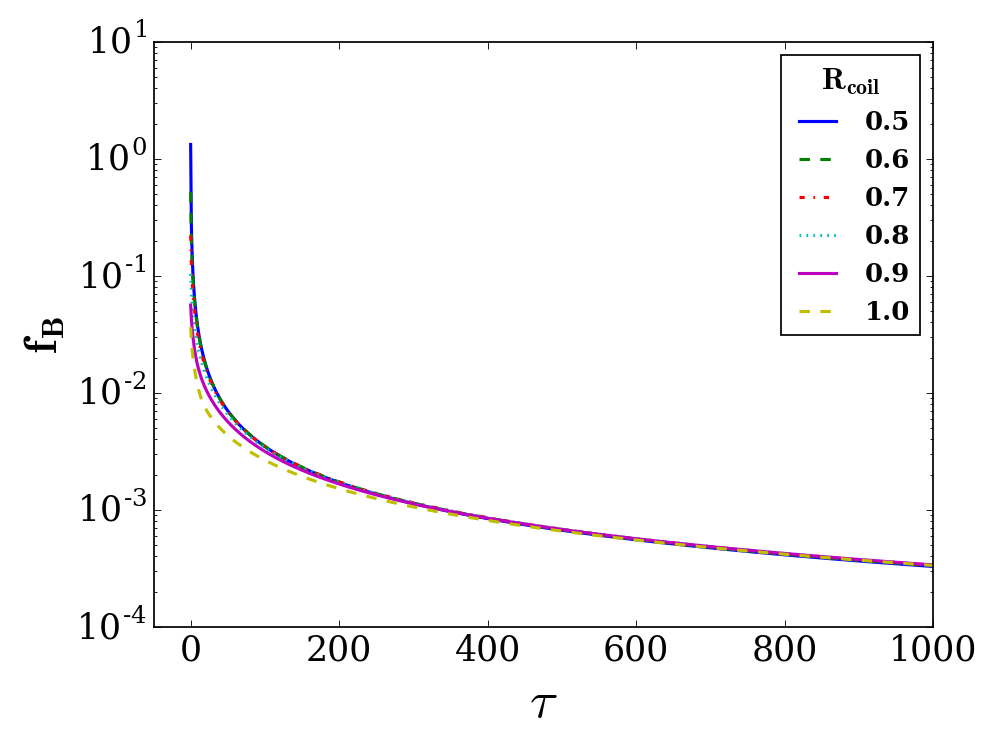}}
 \subfigure[] {\label{fig:wLconv}\includegraphics[width=0.48\textwidth]{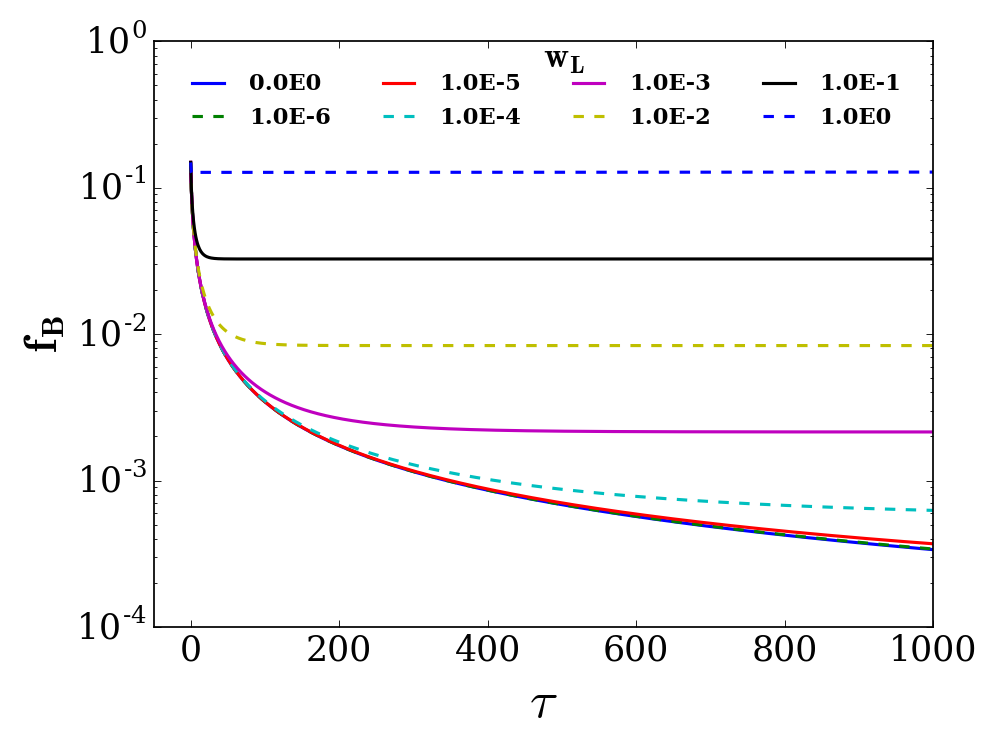}}
 \caption{Convergence explorations of the FOCUS code. (a) Varying the number of Fourier harmonics representing the coils. (b) Changing the total number of initial coils. (c) Different radii of initial circular coils. (d) The effect of different weights for the length constraint.\label{fig:convergence}}
 \end{figure}

\begin{figure}[!htb]
 \subfigure[] {\label{fig:l2_NF1}\includegraphics[width=0.32\textwidth]{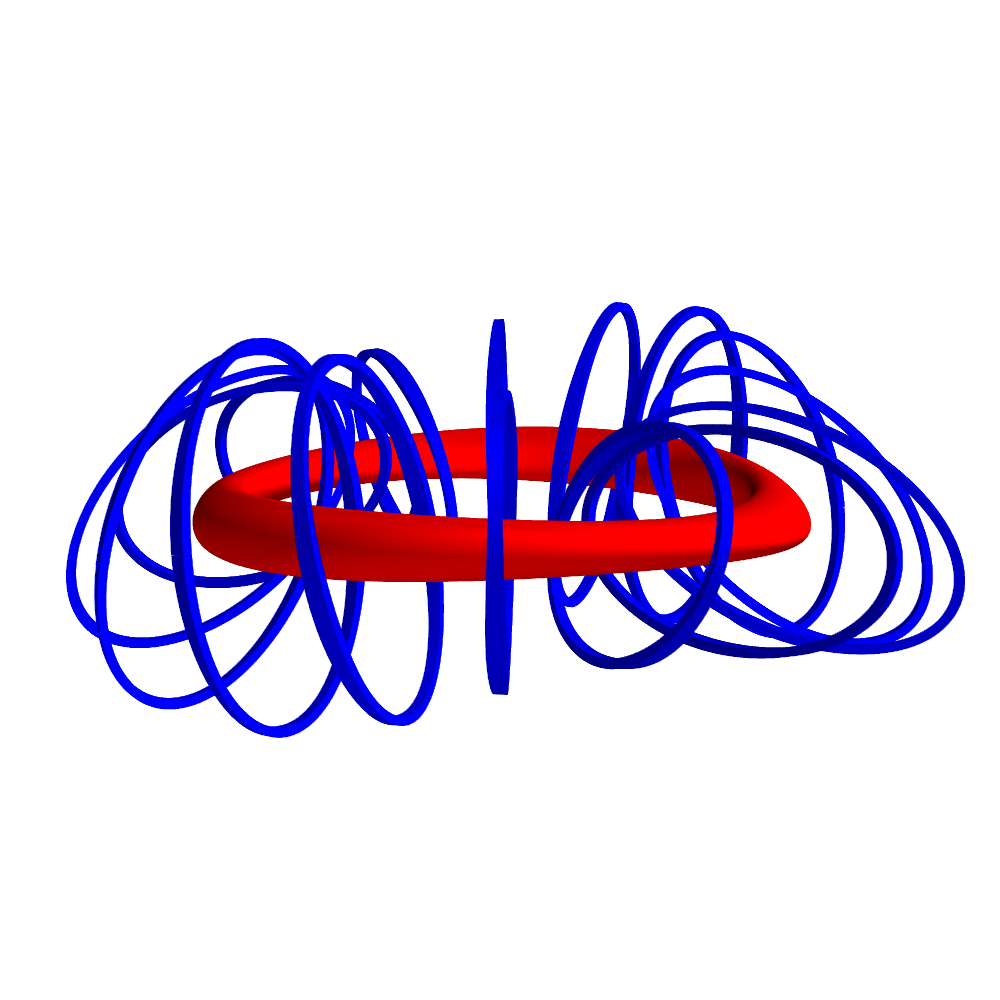}}
 \subfigure[] {\label{fig:Nc4}\includegraphics[width=0.32\textwidth]{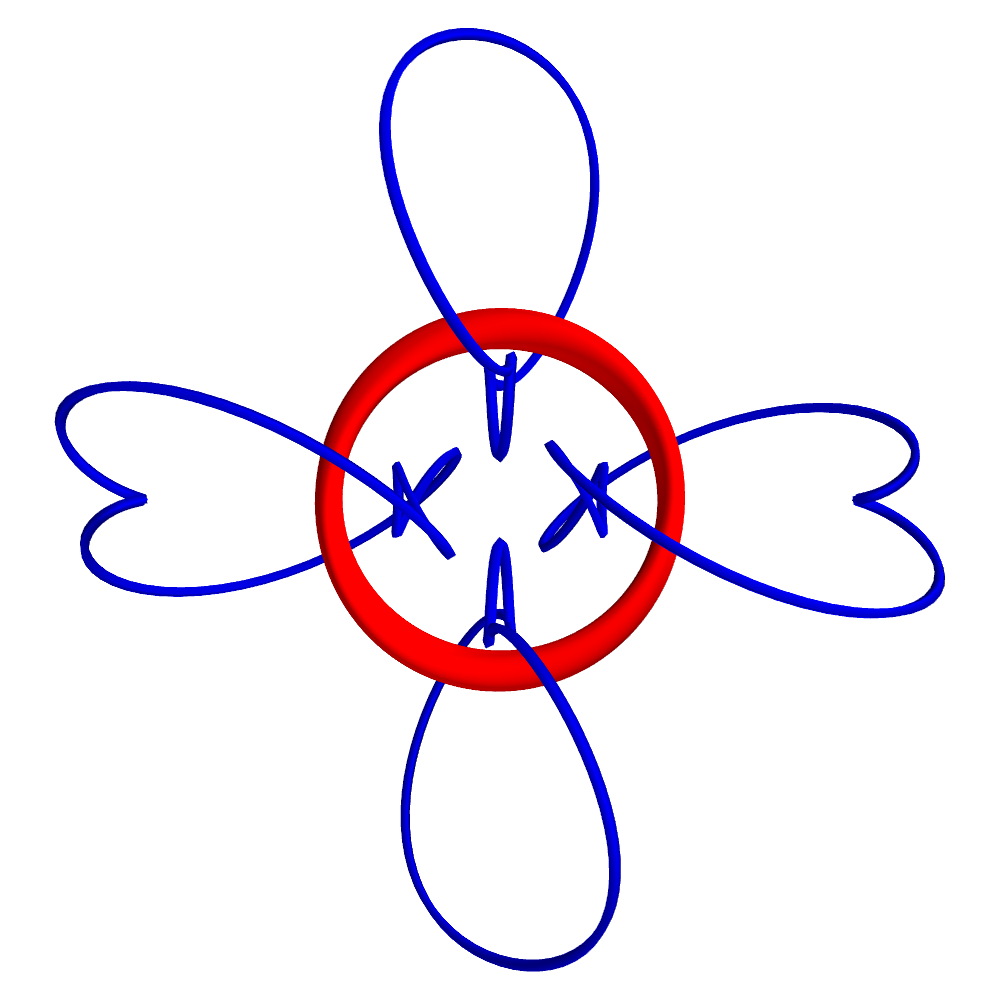}}
 \subfigure[] {\label{fig:L2final2}\includegraphics[width=0.32\textwidth]{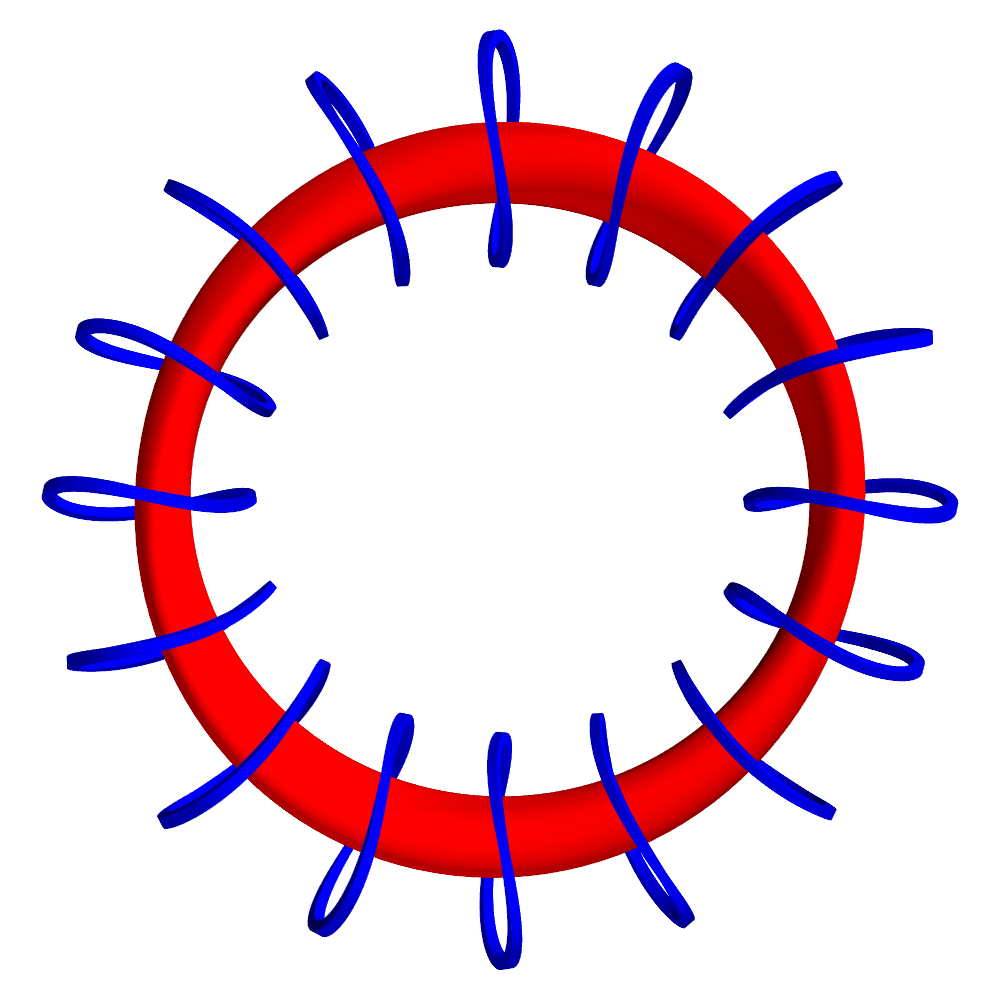}}
 \caption{Different coil shapes from FOCUS optimizations. (a) Solution of only planar coils ($N_F=1$). (b) With only four coils($N_C=4$). (c) Satisfactory coils balancing both physical accuracy and engineering simplicity.\label{fig:L2_coils}}
 \end{figure}

We now vary the Fourier harmonics number, $N_F$, to explore its relations with the achieved normal field error.
As shown in \Fig{NFconv}, when the number of Fourier harmonics is restricted to $N_F=1$, which means that the coils can only be planar, the normal field error cannot decrease as much amplitude as other cases.
The optimized coils, as illustrated in \Fig{l2_NF1}, are elliptical and rotated to follow the pattern of the plasma shape.
And for $N_F\geq 4$ cases, there are no distinct differences.
In other words, for this elliptical configuration, the coils can be adequately represented by $N_F = 4$.
 
The number of initial coils, $N_C$, can also be changed.
In principle, the more coils, the more accurate magnetic field we can get.
However, more coils imply higher cost and less separations between each coils which are vital for diagnostic ports.
Usually, we have to determine the most balanced number of coils for a practical machine.
\Fig{Ncconv} shows the convergence curves with respect to different number of coils.
As expected, with more coils, the lower $f_B$ has been achieved.
Also, we observe that there exists a big gap between the curves $N_C = 32$ and $N_C = 36$,  which implies that this could be the most effective number.
Without enough coils, the shape of the finally optimized coils would be unrealistically complicated, like the coils in \Fig{Nc4}.
The four coils have large twists to produce the desired magnetic field.

For the calculations shown in \Fig{NFconv} and \Fig{Ncconv}, the initial coils are circular coils with a arbitrarily chosen radius of $R_C = 0.75m$.
For different initial radii (of course differing in a small range), as shown in \Fig{Rconv}, we found that the final $B_n$ error are trending to be identical.
And all the optimized coils are having close geometries (details not shown here), which indicates that they are at the same minima.
So, the optimizing process is not very sensitive to the initial radius, at least for this equilibrium.

Now, we explore the effect of length constraint weight.
We assign $w_L$ with different values ranging from $10^{-6}$ to $1.0$. 
The results in \Fig{wLconv} show that the constraint on coil length are suppressing the decrease of the normal field error.
And the evolutions now converge faster as the length weight increases.
This can be used to control the convergence of the code and make a balance between the physical accuracy and engineering feasibility.
 
 With the experience of the previous explorations, we can come up with a suitable coils set for the rotating ellipse configuration by enabling the length constraint.
 All the three constraints are activated with normalized weights so that at the initial state $w_Bf_B = 1.0$, $w_{\Psi}f_\Psi = 0.5$ and $w_L f_L = 0.2$.
 Starting with 16 circular coils with a radius of $0.75 m$, we get a satisfactory coils set at $\tau=10^3$ as shown in \Fig{L2final2}, which makes a reasonable tradeoff between all the constraints. The coils shape now looks much more simple with a comparably smaller size.
 
 \subsection{Designing modular coils for the W7-X} 
 The rotating ellipse is just a simple fictitious classical stellarator configuration.
 It would be more challenging to design coils for the W7-X, due to its complicated geometries.
 The coils system of W7-X mainly consists 50 non-planar modular coils and 20 planar coils in five-field periods \cite{w7xcoils}.
 Modular coils produce the required magnetic field, while the planar coils are mainly used for the flexibility of different rotational transform ($\iota$) scenarios.
 The last closed flux surface (LCFS) in the ``standard case'' configuration \cite{w7x_standard} is used here as the target plasma boundary.
 To produce this configuration, the planar coils are not used (zero currents) and the currents in modular coils are $1.62$ MA (108 turns $\times$ 15.0 kA/turn).
 
\begin{figure}[h]
 \centering
 \begin{minipage}[b]{0.48\textwidth}
 \includegraphics[width=\textwidth]{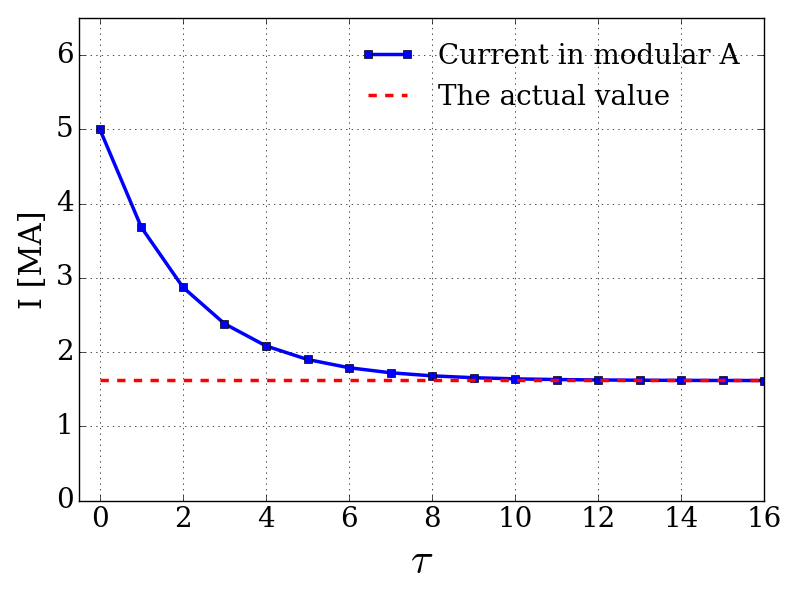}
 \caption{Optimizing the current in one single W7-X coil from a incorrect starting value. \label{fig:w7x_perturbedI}}
  \end{minipage}
  \hfill
  \begin{minipage}[b]{0.48\textwidth}
 \includegraphics[width=\textwidth]{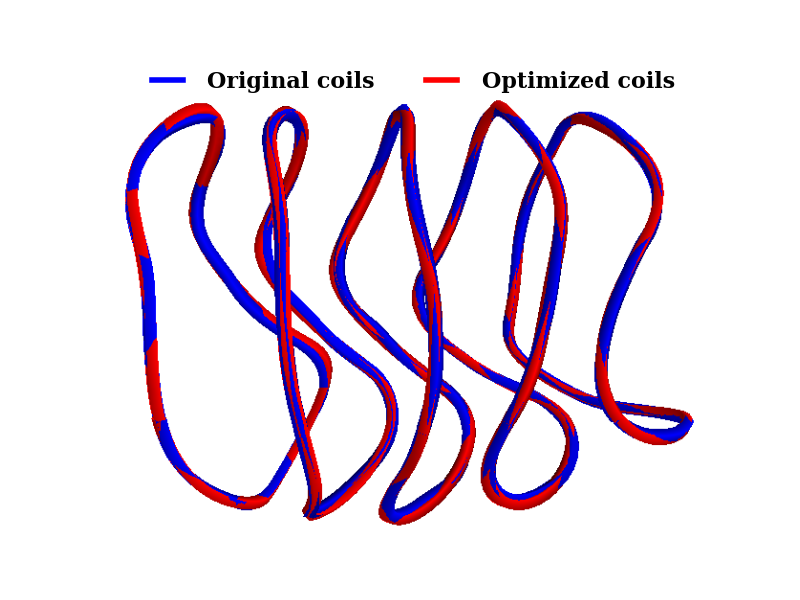}
 \caption{Optimized coils with lower $B_n$ error, starting from the actual W7-X coils. Differences between the two coils sets are indistinguishable at this scale.\label{fig:w7x_further_coils}}
 \end{minipage}
 \end{figure}

 First of all, some preliminary calculations are performed to verify FOCUS' availability for the W7-X.
 The actual W7-X coils (single-filamentary CAD models of the as-built W7-X coils set) are fitted with $N_F=6$ Fourier series and taken as the initial coils.
 One of the simplest verifications is perturbing only one parameter and remaining all the others unchanged to see if FOCUS can recover the perturbed parameter
 (Here we are assuming that the actual coils are already at/close the minimum.)
 The current in an arbitrary modular coil (marked as modular A for notation) is changed to $5.0$ MA, and as shown in \Fig{w7x_perturbedI}, FOCUS can recover the current to the correct value of $1.62$ MA.
 Another verification is taken out by optimizing all the $2000$ parameters of the actual 50 modular coils to lower $B_n$ error.
 The original $f_B$ is $1.86 \times 10^{-4}$. 
 And after a quick FOCUS optimization, the $f_B$ is reduced to $3.56\times 10^{-5}$.
 Although, the differences between the optimized coils and the original ones are visually subtle, as shown in \Fig{w7x_further_coils}.

 Now, we'll start the optimization processes with an arbitrary initial guess to better illustrate the capability of FOCUS.
 The initial guess comes from a set of circular coils ($R_{coil}=1.25$m) that are positioned surrounding the plasma surface with equal toroidally-interval angles, i.e., the $i^{th}$ coil is placed in the poloidal plane of \mbox{$\zeta = 2\pi(i-1)/{N_C}$}.
 The total number of coils ${N_C}$ varies from 50 down to 20 to obtain various solutions.
 All the three constraints introduced above are taken into account with artificially chosen weights.
 
\begin{figure}[!htb]
 \centering
 \includegraphics[width=\textwidth]{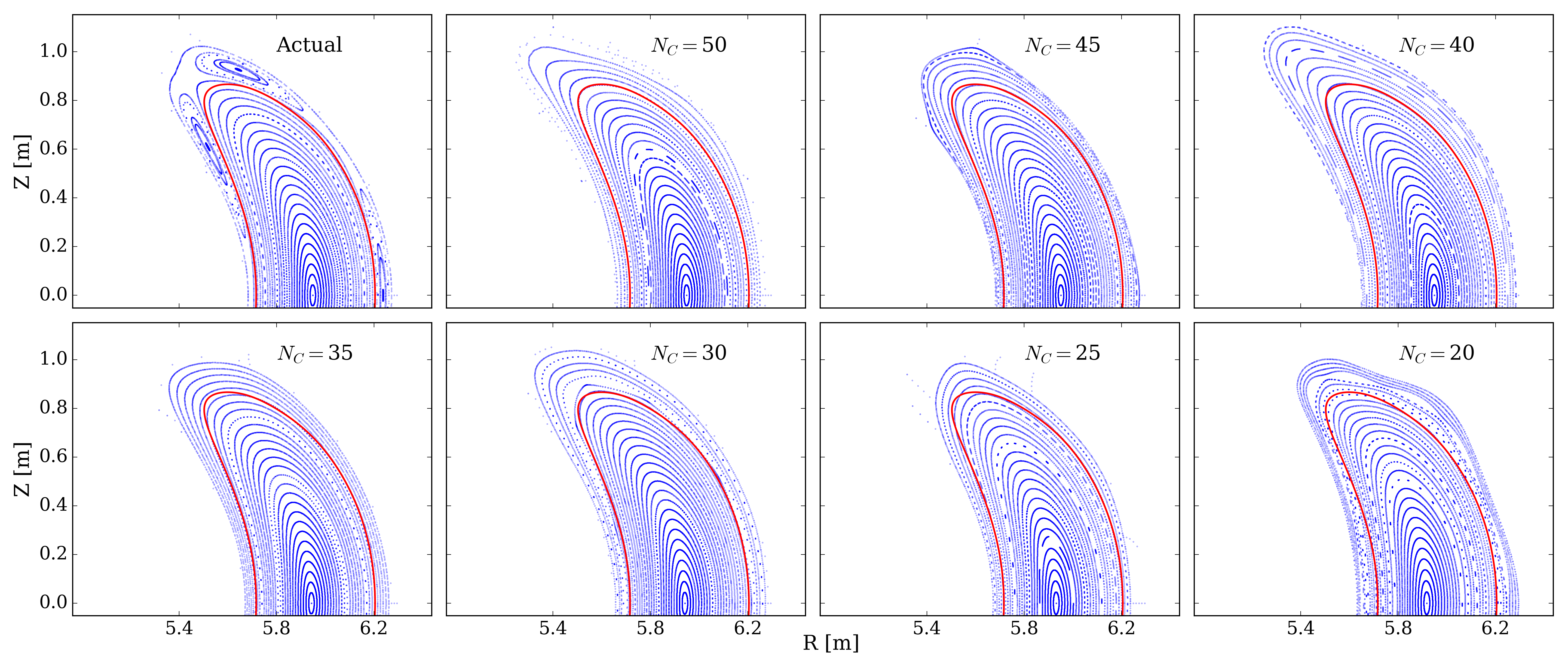}
 \caption{\Poincare plots of the flux surfaces at $\z =0$ bean-shaped plane produced by the actual coils (leftmost in the first row) and FOCUS optimized coils (the others). All the FOCUS coils are optimized from toroidally equally-spaced circular coils with a total number of $N_C$. The target plasma boundary (red curves) is the LCFS of the standard configuration. Only the upper halves are shown here due to the up-down symmetry. \label{fig:w7xpoincare}}
 \end{figure}

 After the optimizations, \Poincare plots of the flux surfaces produced by the magnetic field of the optimized coils, as well as the actual coils, are produced by field-line tracing, as shown in \Fig{w7xpoincare}.
It's clear that FOCUS can produce reasonably close agreements at the target plasma boundary, even with only 20 coils.
The main visible difference is that the $n/m=5/5$ island chain outside the LCFS in the standard configuration is not visual in the \Poincare plots of FOCUS coils.
The magnetic island content is determined by the rotational-transform profile and the magnetic perturbation spectrum.
There are clearly small differences in both between the FOCUS coils and the actual W7-X coils; however, this is to be expected.
As the number of coils used in FOCUS is reduced, the ability to exactly reconstruct a given boundary is also reduced.

The ``magnetic perturbation'' as defined in this paper is ${\bf B}\cdot{\bf n}$ on the given reference boundary, and it is this quantity that is minimized during the optimization.
However, realizing that it is only the resonant part of this perturbation that is associated with the formation of magnetic islands, in future work we intend to (i) identify the resonant harmonics of the perturbation spectrum that are more important to control; and (ii) include additional weights in our target function that explicitly penalize the resonant harmonics of the perturbation spectrum.
We expect that this will allow greater control of the island content.
 
\begin{figure}[!htb]
 \centering
 \includegraphics[width=\textwidth]{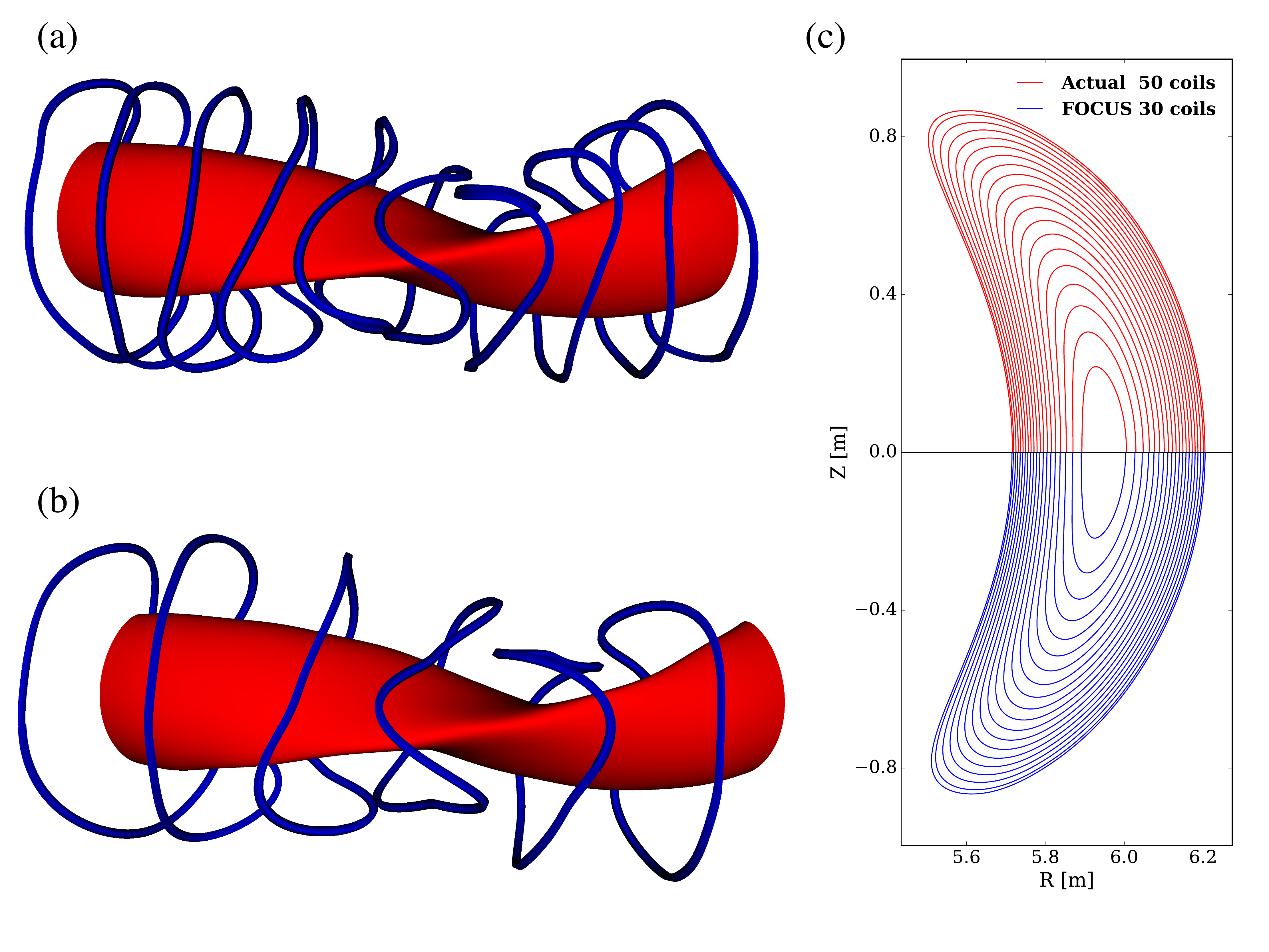}
 \caption{Comparison of the actual W7-X coils (a) and the $N_C=30$ FOCUS coils (b) with their VMEC constructed plasma boundaries (c). In the left-upper figure, one period of the 10 actual W7-X coils (blue) are plotted with its produced plasma surface (red). The left-lower figure shows the 6 modular coils (blue) in the same period from the FOCUS optimizations. And on the right, the nested flux surfaces produced by the actual coils set (red, upper half) and the $N_C=30$ FOCUS coils (blue, lower half) with VMEC free-boundary calculations are presented.}
 \label{fig:focus30}
 \end{figure}

 This paper is primarily concerned with the introduction of a new coil optimization algorithm, and the W7-X calculation is used for an initial illustration.
 We have not performed a comprehensive experimental design study of W7-X and we do not wish to elaborate upon this calculation; however, there is one calculation we present that, in a practical sense, quantifies the difference between the actual W7-X coils and the coils set presented above.
 Ultimately, the magnetic field provided by the coils is required to confine a plasma equilibrium.
 So, we compute the free-boundary,  MHD equilibrium using the widely used VMEC \cite{VMECfb} code with zero pressure and zero plasma currents.
 \Fig{focus30} shows the differences in the nested flux surfaces of the $N_C=30$ coils set and the actual coils, together with a comparison of the coil shapes.
 The results indicate that the two coils sets have comparatively close plasma boundaries.

\subsection{Designing helical coils for the LHD}
\begin{figure}[!htb]
 \centering
 \includegraphics[width=0.5\textwidth]{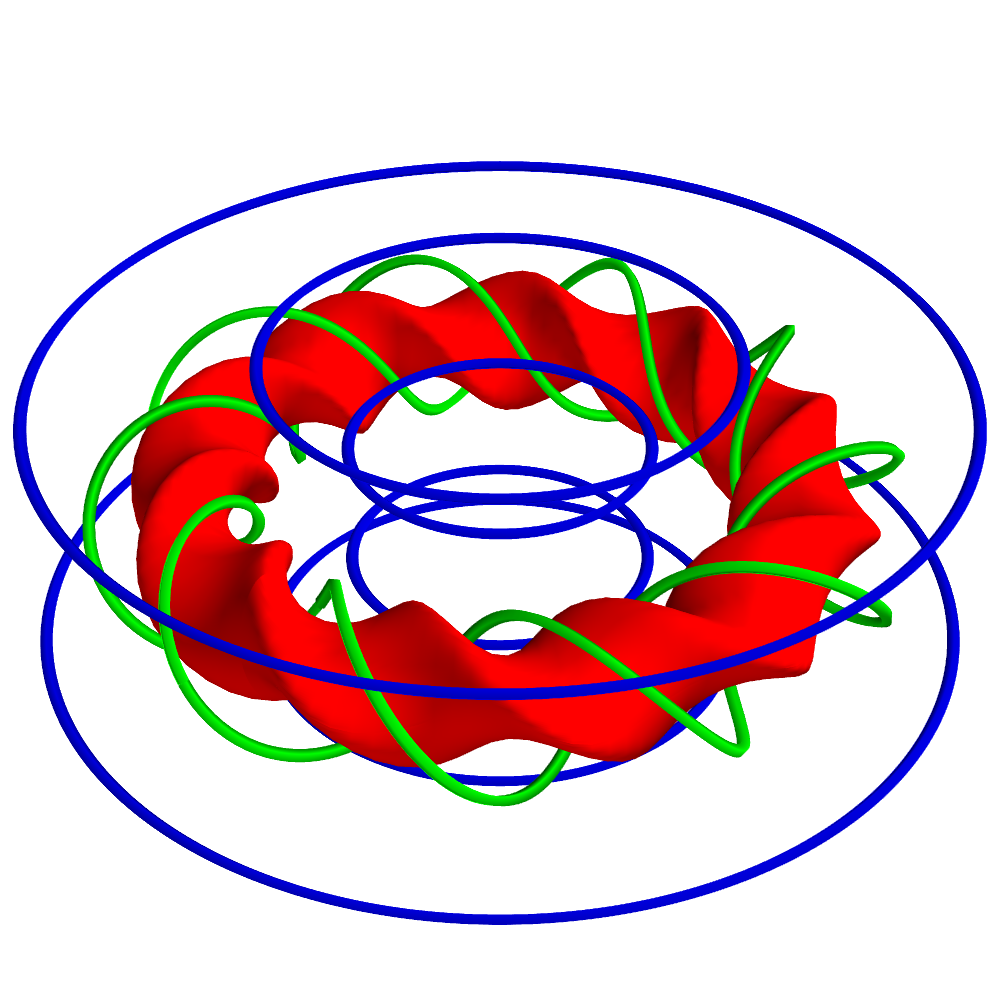}
 \caption{The LHD coils (simplified as filaments) consist of two helical windings (green) and six vertical coils (blue).  \label{fig:LHD_coils}}
 \end{figure}

Besides the W7-X device, another existing large stellarator/heliotron machine is the Large Helical Device (LHD) \cite{LHDdesign}, in Japan. 
Unlike the modular coils used in W7-X, the main coils of LHD are a pair of helical windings.
The approximated LHD coils, including two helical windings and three pairs of vertical field coils,  are shown in \Fig{LHD_coils}. 
Helical curves on a torus can be represented by the Fourier series precisely:
\begin{align}
x & = [ R + r \cos(\theta) ] \cos(N\theta) \ ;\nonumber \\
y & = [ R + r \cos(\theta) ] \sin(N\theta) \ ; \\
z & = r \sin(\theta) \nonumber \ ;
\end{align}
where $R$, $r$ are the major, minor radius of the torus and $N=10/2$ for the LHD case.

\begin{figure}[!htb]
 \centering
 \includegraphics[width=0.8\textwidth]{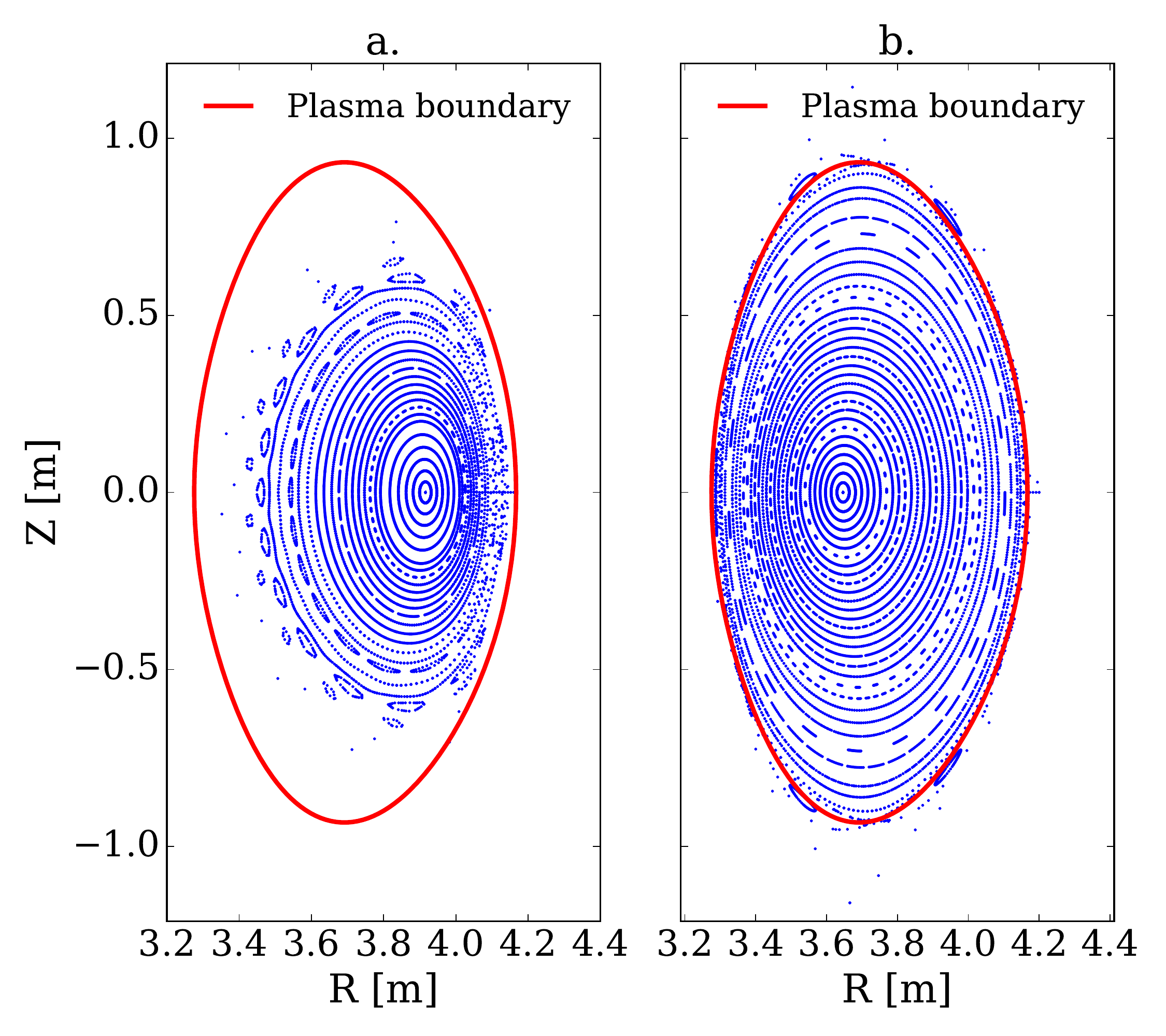}
 \caption{\Poincare plots of the flux surfaces produced by the initial (a) and optimized (b) coils of LHD at $\zeta = 0$ plane, compared with the target plasma boundary (black).\label{fig:LHD_pp}}
 \end{figure}

When fitting the actual LHD coils with $N_F=6$ Fourier series, truncating errors are introduced.
Consequently, the fitted coils have slightly different geometries from the actual coils, which ultimately results in the relatively large errors for the produced magnetic field ($f_B= 1.062\times 10^{-1}$).
\Fig{LHD_pp} (a) shows that the initial fitted coils have deteriorate magnetic flux surfaces.
There are no closed flux surfaces at the target boundary and the magnetic axis is shifted outwards badly.
When we use FOCUS to optimize the fitted coils,  the ${\bf B}\cdot{\bf n}$ error $f_B$ is decreased to $7.112\times 10^{-3}$ at at $\tau=10^2$.
The \Poincare plots in \Fig{LHD_pp}(b) indicate that the optimized coils fix the errors and can produce remarkably good flux surfaces that match the target plasma boundary.

 \section{Conclusions}
 In this paper, we have described a new method for designing stellarator coils without the winding surface and introduced the numerical tool, the FOCUS code.
 
 Theoretical interpretations show that the winding surface is not essential. 
 Fourier series is chosen to represent the three-dimensional coils. 
 And for the first time, we have analytically derive the relations between coil parameters and the target functions. 
 This offers a better physical guidance for how to minimize the target function and is then used to analytically calculate the first derivatives.
 The steepest descent minimization algorithm is successfully employed to optimize a well-constructed target function that includes both physical requirements and engineering constraints.
 Numerical applications have been performed.
 The convergence studies show that the FOCUS code is quite effective and flexible in finding appropriate coils.
 Results of the W7-X and LHD cases indicate that the code is also applicable to practical complicated configurations, with both modular coils and helical coils.
 And by getting rid of the winding surface, the FOCUS code has the potential to find out more feasible coil solutions.
 
 In the future, we will keep constructing more constraints regarding to different needs, like optimizing the resonant perturbation harmonics and other engineering constraints.
 And also, we intend to implement more advanced algorithms.
 The steepest descent is not usually the fastest optimization method, and it's a local minimizer, but it is used here for simplicity of illustration.
 The capacity of handling configurations that have nonzero plasma currents will also be carried out, such that the code can be used for more general cases, even for designing 3D coils in tokamaks.

 \section{Acknowledgements}
 This work was supported by the Chinese Scholarship Council (CSC) with No. 201506340040 and U.S. Department of Energy with under DE-AC02-09CH11466.
 The authors gracefully acknowledge the fruitful discussions with S. Lazerson, J. Breslau, D. Gates and N. Pomphrey.
 Special appreciations are expressed to J. Geiger and J. Loizu from IPP for providing the W7-X data, and Y. Suzuki from NIFS for providing the LHD data.
 The author C.Z. would also like to thank the PPPL Theory Department for hosting his visiting with this work.
 
\section*{References}

\bibliography{focusbib}

\end{document}